\documentstyle[prl,aps,epsf]{revtex}
\begin {document}

\draft
\preprint{ITP-UH-05/98}
\title{\bf Doping of a Spin-1 Chain: Integrable Model}
\author{Holger Frahm$^\dagger$,
	Markus~P.\ Pfannm\"uller$^\dagger$ and
	A.~M.\ Tsvelik$^\ddag$}
\address{
$^\dagger$Institut f\"ur Theoretische Physik, Universit\"at Hannover,
  D-30167 Hannover, Germany}
\address{
$^\ddag$Department of Physics, University of Oxford, 1 Keble Road,
  Oxford, OX1 3NP, UK}

\date{March 1998, revised July 1998}
\maketitle
\begin{abstract}
An exactly soluble model describing a spin $S=1$ antiferromagnetic
chain doped with mobile $S={1/2}$ carriers is constructed. In its
continuum limit the undoped state is described by three gapless
Majorana fermions composing the $SU(2)$ triplet. Doping adds to this a
scalar charge field and a singlet Majorana fermion with different
velocity. We argue that this mode survives when the Haldane gap is
added. 
\end{abstract}
\pacs{71.10.Pm 
      75.10.Jm 
      75.50.Ee 
     }

%
It is well known that mobile holes introduced into antiferromagnetic
Mott insulators by means of doping cause frustration which seriously
affects the magnetic properties of the system. In this paper we study
a doped spin-1 chain which in the presence of mobile holes
interpolates between $S=1$ and $S=1/2$ states.  According to
Refs.~\onlinecite{Penc95,dagx:96,riera:97} this situation is realized
in the carrier doped Haldane system Y$_{2-x}$Ca$_x$BaNiO$_5$
\cite{DiTusa94}: For $x=0$ the interactions between the spin-1
Ni$^{2+}$ ions are well described by a Heisenberg model, upon doping
mixing of the $S=1/2$ holes on the Oxygen sites leads to a low energy
doublet state in an effective one-band Hamiltonian which can move in
the $S=1$ background.

The model  we study is an  integrable model whose Hamiltonian
closely resembles the Hamiltonian introduced in \onlinecite{dagx:96}:
\begin{eqnarray}
   {\cal H} &=& \sum_{n=1}^L \left\{
	{\cal H}^{\rm exch}_{n,n+1} + {\cal H}^{\rm hopp}_{n,n+1}\right\}\ ,
\label{hamil}\\
   {\cal H}^{\rm exch}_{ij} &=&
        {1\over2}\left(
	 {1\over S_i S_j} {\mathbf S}_i \cdot {\mathbf S}_j - 1
	 +\delta_{S_iS_j,1}\left(1-({\mathbf S}_i\cdot{\mathbf S}_j)^2\right)
	\right)\ ,
\nonumber\\
   {\cal H}^{\rm hopp}_{ij} &=& 
	-\left(1-\delta_{S_i,S_j}\right)
	 {\cal P}_{ij} \left( {\mathbf S}_i \cdot {\mathbf S}_j \right)\ .
\nonumber
\end{eqnarray}
Here ${\mathbf S}_i^2=S_i(S_i+1)$ with $S_i=1$ or ${1/2}$ and
${\cal P}_{ij}$ permutes the spins on sites $i$ and $j$.  Comparing
(\ref{hamil}) to the effective one-band Hamiltonian for the doped Ni
oxide given in \onlinecite{dagx:96},
\begin{equation}
 {\cal H}_{ij} =
	\delta_{S_iS_{j},1}
		J{\mathbf S}_i\cdot{\mathbf S}_{j + 1}
	- {\cal P}_{ij} \left( {\mathbf S}_i\cdot{\mathbf S}_{j} +
	   {1\over2} \right)
\label{hdag}
\end{equation}
we find several differences. First, the spin exchange between $S=1$
sites contains biquadratic terms giving the spin-1 Takhtajan-Babujian
chain for hole concentration $x=0$. Therefore in the undoped limit the
spectrum is gapless and one may think that 
the most spectacular feature of the doped $S =
1$ chain -- filling the gap with holes, is lost in our model. We shall
see later however, that it is possible to reintroduce a gap in the
continuous limit where we have a field theoretical 
description of the model.  Second, due to a
hidden supersymmetry of the model (see below) the ratio between the
exchange integral and the hopping is fixed in (\ref{hamil}).  We do
not expect this to be a serious restriction on physical properties of
the model.  Third, hopping processes such as $|1_i\downarrow_j\rangle
\to |\downarrow_i 1_j\rangle$ are allowed.  Finally the integrable
model contains an antiferromagnetic exchange interaction between
nearest neighbour hole states which is responsible for the absence of
a ferromagnetic phase of (\ref{hdag}) for
sufficiently large hole concentration.  Such an interaction has,
however, been considered in Ref.~\onlinecite{riera:97} to improve
agreement with the experiments at $x=1$.

In the following we will use the exact solution of (\ref{hamil}) to
deduce an effective field theory for the low-energy sector of the
system which will enable us to study possible
relevant perturbations.  The low-energy limit of the undoped system
($x=0$) is well known to be a realization of an $SU(2)$ level $k=2$
Wess-Zumino-Novikov-Witten (WZNW) model with central charge $c=3/2$.
This model is equivalent to a model of three massless Majorana
fermions composing an $SU(2)$ triplet \cite{zafa:86}.  A completely
filled ($x=1$) band corresponds to the $S = 1/2$ chain which in the
low-energy limit is equivalent to the $SU(2)_1$ WZNW model.  For
finite doping we find, as may be expected, one free bosonic mode in
the charge sector.  The spin sector however, turns out to be rather
unusual containing a direct sum of $c=3/2$ and $c=1/2$ models with
different velocities. Thus doping generates the fourth Majorana
fermion -- a feature observed in various models related to two-channel
Kondo physics \cite{Kondo,coleman}.

The model (\ref{hamil}) is constructed from solutions to the Yang
Baxter (YBE) equation invariant under the action of the graded Lie
algebra $gl(2|1)$ in the `atypical' representation $[S]_+$
\cite{gl21}.  These representations contain two multiplets of spin $S$
and $(S-{1/2})$ with respect to the $SU(2)$-subalgebra of $gl(2|1)$.
A well known example for a lattice model obtained from the
$\left[{1/2}\right]_+$-representation in this approach is the
supersymmetric $t$--$J$ model (see e.g.\ \onlinecite{susyTJ}).
Denoting the generators of the $SU(2)$-subalgebra by $S^a$, the
$U(1)$-charge operator by $B$ and the remaining fermionic generators
of $gl(2|1)$ by $V^\pm$, $W^\pm$ (which create and annihilate the
holes, see \onlinecite{gl21}) the local ${\cal L}$-operator of the
Quantum Inverse Scattering Method reads
\begin{equation}
   {\cal L}(\mu) \propto \left( 
	\begin{array}{ccc}
           \mu+ 2iB	& i\sqrt{2}W^-	& i\sqrt{2} W^+ \\
	   i\sqrt{2}V^+	& \mu+i(B+S^z)	& -iS^+ \\
	  -i\sqrt{2}V^-	& -iS^-		& \mu+i(B-S^z)
	\end{array}\right) .
\label{lop}
\end{equation}
The spectrum of the transfer matrix for a vertex model with $L$ of
these weights per row is obtained by means of the algebraic Bethe
Ansatz \cite{vladb}.
Starting from the fully polarized state of spin-$S$ multiplets on each
site we consider states with $N_h$ holes (generating sites with spin
$S-{1/2}$) and magnetization $M^z=LS -{1\over2}N_h - N_\downarrow$.
This leads to an auxiliary eigenvalue problem for an inhomogenous
graded six-vertex model on a lattice of $n=N_h+N_\downarrow$ sites
which is solved by means of a second Bethe Ansatz through addition of
states with fermionic grading (holes) to its eigenstate with
$n=N_\downarrow$ (for details see e.g.\ \onlinecite{susyTJ}).  As a
result the spectrum of the supersymmetric vertex model is parametrized
by the roots of the following set of Bethe Ansatz equations (BAE)
\begin{eqnarray}
   \left( {\lambda_j+iS\over \lambda_j-iS} \right)^L &=&
       \prod_{k\ne j}^{N_h+N_\downarrow}
	{{\lambda_j-\lambda_k+i}\over{\lambda_j-\lambda_k-i}}\,
       \prod_{\alpha=1}^{N_h}
	{{\lambda_j-\nu_\alpha-{i\over2}} \over
	 {\lambda_j-\nu_\alpha+{i\over2}} }\ ,
\nonumber\\
	&& \qquad j=1,\ldots,N_h+N_\downarrow
\label{bae3}\\
   1 &=& \prod_{k=1}^{N_h+N_\downarrow}
	{{\nu_\alpha - \lambda_k +{i\over2}} \over
	 {\nu_\alpha - \lambda_k -{i\over2}}}\ ,
   \quad \alpha=1,\ldots,N_h\ .
\nonumber
\end{eqnarray}
For $S={1/2}$, where the model becomes the supersymmetric $t$--$J$
model, and Eqs.~(\ref{bae3}) are Sutherland's form of the BAE
\cite{suth:75}.  The grading of the underlying algebra allows to
construct two more sets of (equivalent) BAE \cite{kulish:85,susyTJ}.
Eqs.~(\ref{bae3}) are most convenient for our analysis, however.

To construct a \emph{local} Hamiltonian for $S=1$ we project the
graded tensor product ${\cal L}_a(\mu+ {i\over2})
\widehat{\otimes}{\cal L}_b(\mu- {i\over2})$ in two copies $V_a$,
$V_b$ of the matrix space of (\ref{lop}) onto the $[1]_+$
representation in $V_a\otimes V_b$.  The result ${\cal
L}_{[1]_+}(\mu)$ of this fusion \cite{kusk:81,pffr:96} defines a new
vertex model with eigenstates again parametrized by the roots of
(\ref{bae3}).  The logarithmic derivative of its transfer matrix at
$\mu=0$ is the Hamiltonian (\ref{hamil}) with eigenvalues
\begin{eqnarray}
  && E\left(\{\lambda_j\},\{\nu_\alpha\}\right) -HM^z-\tilde\mu N_h
\nonumber\\
  && =  \sum_{k=1}^{N_h+N_\downarrow} \left(H-{2\over\lambda_k^2+1}\right)
     -\sum_{\alpha=1}^{N_h} \left(\mu+{1\over2}H\right) -LH
\label{heig}
\end{eqnarray}
(we have added an external magnetic $H$ field and a (hole) chemical
potential $\tilde\mu=\mu+1$ to the Hamiltonian).  

In the thermodynamic limit the solutions of the BAE (\ref{bae3}) are
real hole rapidities $\nu$, whose density we denote as $\rho(\nu)$ and
complex $\lambda$-strings with densities $\sigma_n(\lambda)$
($n=1,2,\ldots$).  The equations for the densities are
\begin{eqnarray}
    \delta_{n,2}s(x) &=& \sigma_n(x) +
    C_{nm}*\tilde\sigma_m(x) - \delta_{n,1}s*\rho(x)
\nonumber\\[-4pt]
\label{intd}\\[-4pt]
    a_2*s(x) &=& \tilde\rho(x) + [1 + a_2]^{-1}*\rho(x) +
    s*\tilde\sigma_1(x)
\nonumber
\end{eqnarray}
where all densities marked by tilde correspond to holes, $f*g(x)$
denotes a convolution, $s(x) = 1/(2\cosh\pi x)$, $a_n(x) =
(2n/\pi)/(4x^2 + n^2)$ and
\[
   C_{nm}(x) =\delta_{nm} \delta(x) 
	- \left(\delta_{n+1,m}+\delta_{n-1,m}\right)s(x)
\]
Writing the energy (\ref{heig}) in terms of densities of
$\lambda$-holes and $\rho$ we obtain
\begin{eqnarray} 
  E/L &=& -2 -\int{\rm d} x \left[2\pi(a_2*s)+\mu\right]\rho(x)
	  \nonumber\\
      &&  +\int{\rm d} x 2\pi s(x) \tilde\sigma_2(x)
	- \lim_{n\rightarrow\infty}Hn\int {\rm d} x
	   \tilde\sigma_n(x)\ .
\end{eqnarray}

\begin{figure}
\begin{center}
\leavevmode
\epsfxsize=0.45\textwidth
\epsfbox{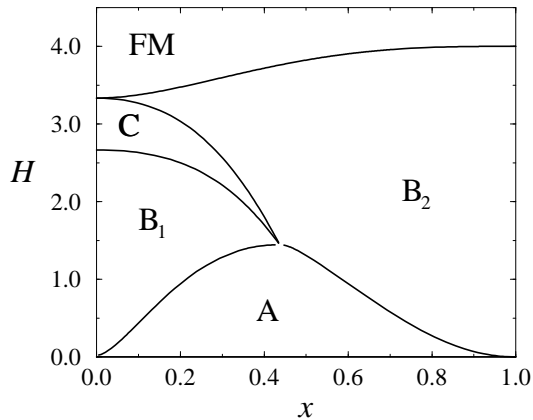}
\end{center}
\caption{Low temperature phase diagram of the doped chain in a
magnetic field $H\gg T$: the saturated ferromagnetic phase is labelled
FM. At intermediate fields the low energy properties of the system are
determined by gapless bosonic modes with energies $\kappa$ (A, B$_1$),
$\epsilon_1$ (A, B$_2$, FM) and $\epsilon_2$ (A, B$_{1,2}$, C).  The
$H=0$ phase is discussed in the main text.
\label{fig:phase}
}
\end{figure}

By minimization the free energy we obtain the thermodynamic Bethe
ansatz (TBA) equations for the energies $\epsilon_n =
T\ln(\tilde\sigma_n/\sigma_n)$ of $\lambda$-strings and $\kappa =
T\ln(\tilde\rho/\rho)$ for the hole rapidities
\begin{eqnarray}
  \epsilon_n(x) &=& Ts*\ln[1 + {\rm e}^{\epsilon_{n -1}(x)/T}]
			  [1 + {\rm e}^{\epsilon_{n +1}(x)/T}] 
\nonumber\\
	&& - 2\pi \delta_{n,2}s(x) 
           - \delta_{n,1}Ts*\ln[1 + {\rm e}^{- \kappa(x)/T}]\ ,
\label{tba1}
\end{eqnarray}
subject to the condition $\lim_{n\to\infty}(\epsilon_n/n) = H$ and
\begin{eqnarray}
  && -[ 2\pi a_2*s(x) + \mu] - Ts*\ln[1 + {\rm e}^{\epsilon_{1}(x)/T}] 
\nonumber\\
  &&\qquad = \kappa(x) + TR*\ln[1 + {\rm e}^{- \kappa(x)/T}]\
\label{kappa}
\end{eqnarray}
where $R = a_2*(1 + a_2)^{-1}$.  In terms of these functions the free
energy reads
\begin{eqnarray}
  F/L &=& -2 - T\int {\rm d} x s(x)\ln[1 + {\rm e}^{\epsilon_2(x)/T}] 
\nonumber\\
      &&\qquad  - T\int {\rm d} x (a_2*s)(x)\ln[1 + {\rm e}^{- \kappa(x)/T}]\ .
\label{FreeE}
\end{eqnarray}
For temperatures $T\ll H$ the energies $\epsilon_{n>2}$ can be
eliminated from (\ref{tba1}), giving the following equation for
$\epsilon_2$
\begin{eqnarray}
  && \epsilon_2(x) -T R*\ln[1 + {\rm e}^{\epsilon_{2}(x)/T}]
\nonumber\\
  &&\qquad	= -2\pi s(x) +H/2 +
   Ts*\ln[1 + {\rm e}^{\epsilon_{1}(x)/T}]
\label{tbaT}
\end{eqnarray}
Together with the $n=1$ equation from (\ref{tba1}) and (\ref{kappa})
this equation determines the low temperature phase diagram of the
system (Fig.~\ref{fig:phase}).  For sufficiently large negative $\mu$
Eq.~(\ref{kappa}) implies $\kappa>0$ corresponding to vanishing hole
density.  The remaining TBA coincide with those for the integrable
$S=1$ magnet \cite{takh:82,Babu}.  Choosing $\mu>H/2$ we find
$\kappa<0$ and $\epsilon_1<0$ which implies a hole density $x=1$.  The
resulting TBA for $\epsilon_{n\ge2}$ are those of the $S={1/2}$
Heisenberg chain.  Similar considerations for intermediate $\mu$ yield
the complete phase diagram and lead us to identify $\kappa$ and
$\epsilon_1$ as the energy of the charge and spin modes associated
with the mobile carriers while $\epsilon_2$ is the magnetic mode of
the background spins.

We now concentrate on the case of finite doping in a vanishing
magnetic field which corresponds to chemical potentials $\pi-4<\mu<0$.
In this regime we have $\kappa < 0$ for $|x| < Q$.  For temperatures
smaller than the Fermi energy of holes $T \ll -\mu$ one can replace
$\kappa$ in Eq.~(\ref{tba1}) by its zero temperature value
$\kappa_0(x)$ and the free terms by their asymptotics. As a result we
get
\begin{eqnarray}
  \epsilon_n(x) &=& Ts*\ln[1 + {\rm e}^{\epsilon_{n -1}(x)/T}][1 +
  {\rm e}^{\epsilon_{n +1}(x)/T}]
\nonumber\\
  &&\quad - 2\pi\delta_{n,2}{\rm e}^{- \pi|x|} 
  - 2\pi A\delta_{n,1}{\rm e}^{- \pi|x|}
\label{tba2}
\end{eqnarray}
where $A = - (2\pi)^{-1}\int_{-Q}^Q{\rm d} y {\rm e}^{\pi
y}\kappa_0(y)$.  To study the specific thermodynamic properties of the
present model we have to separate the contributions to the free energy
stemming from the charge-sector from those due to the $\epsilon_n$.
Considering low temperatures again the leading contributions to
$\kappa$ come from the vicinity of the Fermi wave vectors $\pm Q$. In
this region one can safely neglect contributions to Eq.~(\ref{kappa})
from $\epsilon_1$ and rewrite it as
\begin{eqnarray}
 && -[2\pi a_2*s(x) + \mu] - TR*\ln[1 + {\rm e}^{- |\kappa(x)|/T}]
\nonumber\\
 &&\qquad = \kappa(x) -  \int_{-Q}^Q {\rm d} yR(x - y)\kappa(y)
\end{eqnarray}
where $Q$ is determined by the condition $\kappa(\pm Q) = 0$. 
Using the procedure introduced by Takahashi \cite{taka:71b}, we can
rewrite the free energy (\ref{FreeE}) as $F/L = E_0/L + f_1 + f_2$
where $E_0$ is the ground state energy and 
\begin{eqnarray}
  f_1 &=& - T\int {\rm d} x \rho_0(x)\ln[1 + {\rm e}^{- |\kappa_0(x)|/T}]
	  \approx- \pi T^2/6v\ ,
\label{Free}\\
  f_2 &=& - T\int {\rm d} x s(x)\ln[1 + {\rm e}^{\epsilon_2(x)/T}] 
\nonumber\\
 &&- T\int {\rm d} x (s*\rho_0)(x)\ln[1 + {\rm e}^{\epsilon_1(x)/T}]\ .
\label{free2}
\end{eqnarray}
Here $\rho_0(x)$ is the density of hole rapidities at $T=0$
\[
  \rho_0(x) - \int_{-Q}^Q {\rm d} yR(x - y)\rho_0(y) = a_2*s(x)
\]
and $v$ is the velocity of the charge mode $\kappa$
\begin{equation}
  v = \frac{1}{2\pi\rho_0(Q)}
	\left.\frac{\partial\kappa_0}{\partial x}\right|_{x = Q}\ .
\label{v}
\end{equation}

In the low-T limit $T \ll -\mu$ where this form is valid one may
replace $s(x)$ and $s*\rho_0(x)$ in (\ref{free2}) by their
asymptotics.  Now we have TBA equations in such form that the spin
sector is manifestly decoupled from the charge one.  The free energy
of the latter is given by Eq.~(\ref{Free}) representing a scalar
bosonic mode, while Eqs.~(\ref{tba2}) and (\ref{free2}) describe the
thermodynamics of the spin sector.  Such TBA equations (i.e.\ with two
driving terms) arise also in other systems combining different spins
\cite{Spins} and in the two-channel Kondo model with a channel
anisotropy \cite{Andrei}.  As shown in Ref.~\onlinecite{coleman} such
two-channel Kondo model can be written in terms of four Majorana
fermions.  
These considerations lead us to hypothesize that the effective
low-energy theory for the Hamiltonian (\ref{hamil}) at small doping
is:
\begin{eqnarray}
  {\cal H}_{\rm eff} &=& \int{\rm d}x 
	\left[{\cal H} + \bar{\cal H}\right] 
\label{eff}\\
  {\cal H} &=& -\frac{i}{2}\sum_{a = 1}^3v_2\chi_a\partial_x\chi_a  -
		\frac{i}{2}v_1\chi_0\partial_x\chi_0
		+ g\chi_0\chi_1\chi_2\chi_3
\nonumber
\end{eqnarray}
and a similar expression $\bar{\cal H}(\{\bar\chi\})$ for the left
movers.  The fields $\chi_0$, $\chi_a$ are Majorana fermions and the
marginal coupling $g$ between two sectors in (\ref{eff}) is the only
one possible by symmetry (we have omitted marginally irrelevant
interactions between sectors with different chirality).  A magnetic
field couples to the term $i\int \left(\beta \chi_0\chi_a +
\epsilon_{abc}\chi_b\chi_c\right)$ quadratic in the fermions with some
constant $\beta$.

\begin{figure}
\begin{center}
\leavevmode
\epsfxsize=0.45\textwidth
(a)\epsfbox{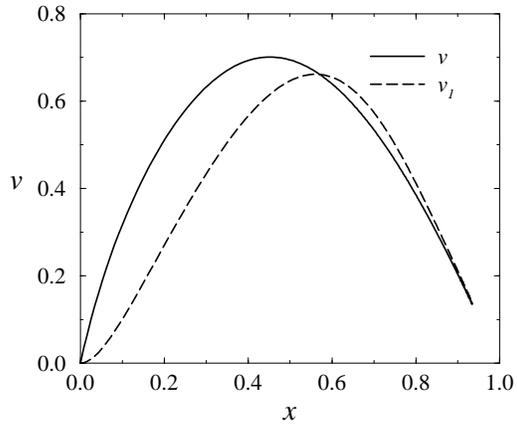}
\epsfxsize=0.45\textwidth
(b)\epsfbox{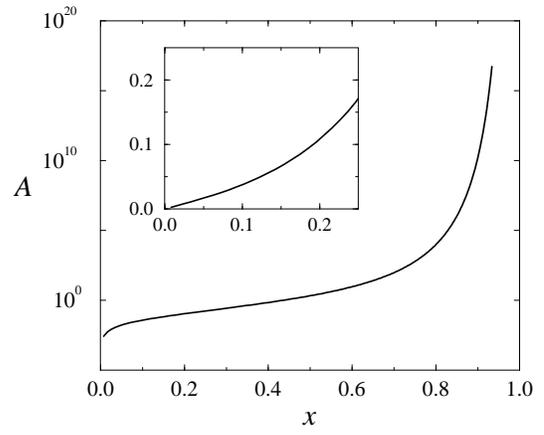}
\end{center}
\caption{Parameters of the effective model (\protect{\ref{eff}}): (a)
velocities $v$ of the charge mode (\protect{\ref{v}}) and $v_1$ of the
fermion $\chi_0$ (\protect{\ref{ratio}}) and (b) the parameter $A$ in
the TBA (\protect{\ref{tba2}}) versus hole concentration $x$. The
inset shows $A$ for small $x$.  Note that $v_2\equiv\pi$ for all
fillings.
\label{fig:velo}}
\end{figure}

The parameters of the effective theory have to be extracted from the
TBA at small doping.  This corresponds to $A\ll1$ and implies that the
two terms in (\ref{free2}) are dominated by contributions from the
region $\pi x\sim \ln(2\pi/T)$ (where $|\epsilon_2| \sim T$) and by
$\pi x \sim \ln(2\pi A/T)$, respectively.  This separation of scales
allows to obtain the small $A$ behaviour of (\ref{free2}) at low
temperatures
\[
  f_2 = -{\pi T^2\over6v_1}\left({1\over2}-{3A\over4\pi}\ln A \right)
	-{\pi T^2\over6v_2}\left({3\over2}+{3A\over4\pi}\ln A \right)
	+ \ldots
\]
For $A=0$ this is the free energy of a single Majorana fermion
($\chi_0$ in (\ref{eff})) with velocity $v_1$ and that of the
$SU(2)_2$ WZNW model with velocity $v_2$ which can be expressed in
terms of massless triplet of Majorana fermions \cite{zafa:86}. The
ratio of the velocities is given by
\begin{equation}
  v_1/v_2 = -{1\over2\pi}\,
	\frac{\int_{-Q}^Q{\rm d} y {\rm e}^{\pi y}\kappa_0(y)}{
		\int_{-Q}^Q{\rm d} y{\rm e}^{\pi y}\rho_0(y)}\ .
\label{ratio}
\end{equation}
The coupling constant $g$ needs to be chosen such that it produces the
$A$-dependence of $f_2$.  Similarly, the parameter $\beta$ can be
determined from the $T = 0$ magnetic susceptibility. At small $A \ll
1$ we get
\[
  \chi = \frac{1}{\pi}\left({1\over v_2}  
	- {1\over v_1}\, {A\over2\pi}\ln{A} +\ldots\right)
\]
giving $\beta^2 = A/2\pi$.
In Fig.~\ref{fig:velo} some of these quantities are given as a
function of the hole concentration $x$.

The simplest $SU(2)$-invariant perturbation of (\ref{eff}) is a mass
term
\(
     i\int {\rm d}x
     \left(m\sum_{a=1}^3\bar\chi_a\chi_a + M\bar\chi_0\chi_0\right)
\)
for the particles in the spin sector (recall that the model of three
massive Majorana fermions is a good continuous approximation for the
$S = 1$ Heisenberg antiferromagnet \cite{tsv1}). Choosing $M=0$ we get
a situation with a Haldane gap in the triplet sector and gapless
singlet excitations.  This may be a good description for the magnetic
states inside the Haldane gap observed in the doped Ni oxide
\cite{DiTusa94}.

To summarize, we have introduced an integrable model describing a
magnetic system which interpolates between the integrable $S=1$
Takhtajan-Babujian and the $S=1/2$ Heisenberg chain.  For finite
doping we find up to three massless modes determining the low
temperature thermodynamics of the system.  In the $SU(2)$ invariant
case ($H=0$) these are a bosonic charge mode and a direct sum of a
$SU(2)_2$ WZNW model and one Majorana fermion with different
velocities in the spin sector.  Further analysis of the TBA is
necessary to gain more insight in the nature of the phase diagram of
this and the related $S>1$ systems.

This work has been supported by the Deutsche Forschungsgemeinschaft under
Grant No.\ Fr~737/2--2.

\setlength{\baselineskip}{14pt}




\begin{thebibliography}{10}


\bibitem{Penc95}
K. Penc and H. Shiba, Phys. Rev. B {\bf 52}, R175 (1995).

\bibitem{dagx:96}
E. Dagotto, J. Riera, A. Sandvik, and A. Moreo, Phys. Rev. Lett. {\bf 76},
  1731  (1996).

\bibitem{riera:97}
J. Riera, K. Hallberg, and E. Dagotto, Phys. Rev. Lett. {\bf 79},
  713 (1997).

\bibitem{DiTusa94}
J.~F. DiTusa {\em et al.}, Phys. Rev. Lett. {\bf 73}, 1857 (1994).

\bibitem{zafa:86}
A. B. Zamolodchikov and V. A. Fateev, Sov. J. Nucl. Phys. {\bf 43}, 
657 (1986).

\bibitem{Kondo}
V.~J. Emery and S. Kivelson, Phys. Rev. B{\bf 46}, 10812 (1992);
%
P. Coleman, L. B. Ioffe and A. M. Tsvelik,  Phys. Rev. B{\bf 52},
  6611 (1995);
%
P. Azaria, C. Hooley, P. Lecheminant, C. Lhuilier and A. M. Tsvelik,
  in preparation. 

\bibitem{coleman} P. Coleman and A. J.  Schofield,
  Phys. Rev. Lett. {\bf 75}, 2184 (1995). 

\bibitem{gl21}
M. Scheunert, W. Nahm, and V. Rittenberg, J. Math. Phys. {\bf 18},  155
  (1977);
M. Marcu, J. Math. Phys. {\bf 21},  1277  (1980).


\bibitem{susyTJ}
F.~H.~L. E{\ss}ler and V.~E. Korepin, Phys. Rev. B {\bf 46},  9147
  (1992);
A. Foerster and M. Karowski, Nucl. Phys. B {\bf 396},  611  (1993).

\bibitem{vladb}
V.~E. Korepin, N.~M. Bogoliubov, and A.~G. Izergin, {\em {Quantum Inverse
  Scattering Method and Correlation Functions}} (Cambridge University Press,
  Cambridge, 1993).

\bibitem{suth:75}
B. Sutherland, Phys. Rev. B {\bf 12},  3795  (1975).

\bibitem{kulish:85}
P.~P. Kulish, J. Sov. Math. {\bf 35},  2648  (1986), [Zap. Nauch. Semin. LOMI
  {\bf 145}, 140 (1985)].

\bibitem{kusk:81}
P.~P. Kulish and E.~K. Sklyanin,  in {\em Integrable Quantum Field Theories},
  Vol.~151 of {\em Lecture Notes in Physics}, edited by J. Hietarinta and C.
  Montonen (Springer Verlag, Berlin, 1982), pp.\ 61--119.

\bibitem{pffr:96}
M.~P. Pfannm{\"u}ller and H. Frahm, Nucl. Phys. B {\bf 479},  575  (1996).

\bibitem{takh:82}
L. Takhtajan, Phys. Lett. A {\bf 87},  479  (1982).

\bibitem{Babu}
H.~M. Babujian, Phys. Lett. A {\bf 90},  479  (1982);
Nucl. Phys. B {\bf 215}, 317 (1983).

\bibitem{taka:71b}
M. Takahashi, Prog. Theor. Phys. {\bf 46},  401  (1971).

\bibitem{Spins}
H.~J. de Vega and F. Woynarovich, J.~Phys. A {\bf 25}, 4499 (1992);
S.~R. Aladim and M.~J. Martins, J.~Phys. A {\bf 26}, L529 (1993);
H.~J. de Vega, L. Mezincescu and R. I. Nepomechie, Phys. Rev. B {\bf 49},
  13223  (1994);
P. Schlottmann, Phys. Rev. B {\bf 49}, 9202 (1994).

\bibitem{Andrei} N. Andrei and A. Jerez, Phys. Rev. Lett. {\bf 74},
  4507 (1995); A. Jerez, N. Andrei and G. Zarand, cond-mat/9803137. 

\bibitem{tsv1} A. M. Tsvelik,  Phys. Rev. B {\bf 42}, 10 499 (1990);
D. G. Shelton, A. A. Nersesyan and A. M. Tsvelik, Phys. Rev. B {\bf
53}, 8521 (1996).   

\end{thebibliography}
\end {document}